\documentclass[12pt]{article}
\usepackage{amsfonts,amsthm,amsmath,amssymb,upgreek}
\usepackage[paper=letterpaper,margin=1in]{geometry}
\usepackage[export]{adjustbox}
\usepackage[numbers,sort&compress]{natbib}
\usepackage{color}
\usepackage[normalem]{ulem}
\usepackage{tikz}
\usepackage{graphicx}
\usepackage{dcolumn}
\usepackage{bm}
\usepackage{hyperref}
\usepackage{float}
\usepackage[caption=false]{subfig}
\usepackage{hyperref}
\hypersetup{
    colorlinks=true,
    linktoc=page,    
    linkcolor=blue,
    urlcolor=magenta
}

\newcommand{\be}{\begin{equation}}
\newcommand{\ee}{\end{equation}}
\newcommand{\bea}{\begin{eqnarray}}
\newcommand{\eea}{\end{eqnarray}}
\newcommand{\dd}{\text{d}}

\newcommand{\bra}[1]{\left<{#1}\right|}
\newcommand{\ket}[1]{\left|{#1}\right>}

\definecolor{revisionblue}{RGB}{0,0,255}
\definecolor{deletionred}{RGB}{220,0,0}

\newcommand{\wkin}{\omega_{\infty}}
\newcommand{\wloc}{\omega_{\mathrm{loc}}}
\newcommand{\Lloc}{\Lambda_{\mathrm{loc}}}
\newcommand{\Iplus}{\mathcal I^{+}}
\newcommand{\Iminus}{\mathcal I^{-}}
\newcommand{\Hplus}{\mathcal H^{+}}

\begin{document}

\thispagestyle{empty}

\vspace*{.5cm}
\begin{center}

{\bf {\Large
Scattering Radiation Emitted Near a Black Hole Horizon}\\
\vspace{1cm}}

\begin{center}

 {\bf Peng Cheng}$^{a}$\footnote{p.cheng.nl@outlook.com}\\
  \bigskip \rm
  
\bigskip
 a) Center for Joint Quantum Studies and Department of Physics, \\School of Science, Tianjin University, Tianjin 300350, China\\

\rm
  \end{center}

\vspace{1.5cm}
{\bf Abstract}
\end{center}
\begin{quotation}

We study a family of outgoing electromagnetic radiation states prepared outside but arbitrarily close to the horizon of a Schwarzschild black hole, with frequencies that remain uniformly bounded in the local source frame.
As the preparation point approaches the horizon, gravitational redshift drives this radiation toward zero Killing frequency and into a controlled soft regime at any fixed exterior scattering radius.
The radiation considered here may be viewed as part of the Hawking radiation spectrum.
After including transmission through the Schwarzschild potential, we obtain the leading soft response in a localized region of weak curvature and formulate an inclusive scattering probability that is infrared safe.
Through the infrared triangle, this response is related to electromagnetic memory at future null infinity.
The conditioned memory record probes angular projections of the zero-frequency component of the transmitted radiation, together with their correlations with the accompanying hard scattering outcome.
It can therefore provide an additional asymptotic constraint and a partial fingerprint of the radiation source.

\end{quotation}

\vspace{1cm}

\setcounter{page}{0}
\setcounter{tocdepth}{2}
\setcounter{footnote}{0}

\newpage
{\noindent} \rule[-10pt]{16.5cm}{0.05em}\\
\tableofcontents
{\noindent} \rule[-10pt]{16.5cm}{0.05em}\\


\section{Introduction}
\label{intro}

It has been proposed that black hole soft hair may constrain aspects of the black hole information paradox (BHIP) \cite{Hawking:2015qqa, Hawking:2016msc, Strominger:2017aeh} (see \cite{Strominger2017} for a review and \cite{Pasterski:2020xvn,Cheng:2020vzw,Cheng:2021gdr, Cheng:2023vrx} for recent developments). 
One version of the BHIP can be summarized as the thermal Hawking radiation \cite{Hawking1975,Hawking:1974rv} cannot carry off all the information of a black hole, and thus the original information encoded inside a black hole seems to be destroyed \cite{Hawking:1976ra}. 
The global charge conservations enforce constraints on the evaporating process (e.g., the total electric charge carried off by Hawking radiation should always equal the loss of the remaining black hole), which might be a way of encoding the information.
However, in classical gravity theory, the number of global charges (i.e., hairs) of a black hole system is too small, which has no ability to encode all the information of the black hole.
The existence of infinite-dimensional asymptotic symmetries in the null infinity region and horizon \cite{Bondi:1962px,Sachs:1962zza} has motivated the association of infinitely many soft charges with a black hole. 
The Hawking-Perry-Strominger (HPS) mechanism states that the conservation law of the extended symmetry can yield infinitely many deterministic constraints on the evaporation process. 
Thus, it provides a way of understanding the information paradox.
Recently, in deriving the Page curve of radiation entropy using soft hair \cite{Cheng:2020vzw}, the role of measurement processes that can decrease the fine-grained entropy is emphasized. 
However, more details of the process, like the physical realization, remain unclear.
These developments motivate an operational question concerning which features of near-horizon radiation survive exterior propagation and remain accessible through asymptotic observables. Soft-decoupling analyses indicate that this information should be characterized through the transmitted state and its correlations with the hard process, rather than attributed to the soft charges alone \cite{Mirbabayi2016,Bousso2017}.
Moreover, the finite-temperature effects on the soft factors and the resulting memory observables have been studied in \cite{Solanki2023}.

The purpose of the current paper is to trace a specified near-horizon radiation family through preparation, exterior propagation, localized scattering, and the asymptotic memory observable.
Let us start with an interesting observation.
Consider one of the classical tests of general relativity, i.e., the Pound-Rebka experiment \cite{Pound1960,Pound1965}, where we have a particle source $A$ located at a position close to a Schwarzschild black hole emitting a particle. 
The particle is redshifted when it is received by the observer $B$ located far from the black hole. 
For a family of emissions whose local source frequency remains bounded, the received frequency tends to zero as the source approaches the infinite-redshift surface.
For Schwarzschild black holes, the infinite redshift surface is the horizon. 
The process can be understood as follows. When the particle moves away from the black hole, it loses energy by overcoming the gravitational force exerted by the black hole.
Our main result begins with the following source condition, which is that outgoing radiation is prepared at $r_e>r_H$ with uniformly bounded local-frequency support, and the limit $r_e\to r_H^+$ is taken only after that state family has been specified.
Moreover, the relation between the Hawking radiation and the particle emitted near the horizon will be discussed in Appendix~\ref{DR}.

In this paper, we determine the controlled soft limit of the specified near-horizon state family after spin-one propagation and study its interaction with a localized hard process at a fixed exterior radius. With the help of the triangular equivalence \cite{Strominger:2013jfa,Strominger:2013lka,Strominger:2014pwa, He:2014laa,Cheng:2022xyr,Cheng:2022xgm,Mao:2023rca,Mao:2023dsy}, we formulate an infrared-safe inclusive probability for the transmitted state and connect the conditioned zero-frequency response with electromagnetic memory at future null infinity. Varying the smearing function $\varepsilon(z,\bar z)$ and conditioning on the hard-scattering record selects the angular projections and correlations that can be compared with features of the prepared state. This provides a concrete framework for tracing how information from the source is carried by the transmitted radiation and ultimately encoded in observables at infinity.

We study the physical effects of the scattering process with the specified near-horizon radiation. The paper is organized as follows.
In section \ref{SoftH}, we define the bounded-local-frequency source family, its observer-dependent frequencies, and its propagation through the electromagnetic greybody channel. 
Section \ref{measure} calculates the scattering amplitude of $n+m$ particles from null infinity and soft radiation from the black hole horizon in momentum space and position space. 
An analysis of the physical consequences of the process is also included in the section. 
Section \ref{Con} summarizes the whole paper and discusses some further issues. The relation between the physical mechanism illustrated in the current paper and Hawking radiation is discussed in Appendix~\ref{DR}. Unless stated otherwise, we use $c=\hbar=k_B=1$ and keep Newton's constant $G$ explicit.

\section{Radiation emitted near a black hole horizon}
\label{SoftH}

This section defines the radiation family used in the main argument before taking any near-horizon limit. We distinguish the conserved Killing frequency from frequencies measured by finite-radius observers, and we propagate the spin-one state through the Schwarzschild exterior before it enters the scattering region. When those particles travel to the exterior region, they would become soft due to the redshift effect. The Damour-Ruffini analysis and its relation to the present source condition are discussed separately in Appendix~\ref{DR}.

\subsection{Near-horizon source state and exterior propagation}

Let $r_H=2GM$ denote the horizon in the areal Schwarzschild coordinate, and define the Schwarzschild lapse in the coordinates used below by
\be
N(x)\equiv\sqrt{-g_{tt}(x)}\,.
\label{lapse}
\ee
We consider outgoing spin-one radiation prepared at a finite radius $r_e>r_H$. Its frequency in the local source frame is required to obey the uniform support condition
\be
0<\wloc(r_e)\leq\Lloc,
\label{boundedLocal}
\ee
where $\Lloc$ is held fixed as $r_e\to r_H^+$. The corresponding conserved Killing frequency is
\be
\wkin=N(r_e)\cdot \wloc(r_e),
\label{sourceFrequency}
\ee
and therefore the entire frequency support of this specified family moves toward $\wkin=0$ in the near-horizon limit. Holding $\wkin$ fixed would define a different family. It is worth noticing that Eq.~\eqref{boundedLocal} is not a statement about the complete Hawking thermal spectrum, whose high-frequency tail is suppressed but not absent \cite{Wald2001,FredenhagenHaag1990}.

A one-photon wave packet is specified on an exterior hypersurface $\Sigma_e$ intersecting the source region at $r_e$
\be
\ket{\Psi_{\mathrm{src}}}
=\sum_{\ell m\lambda}\int\dd\wkin\,
f_{\ell m\lambda}(\wkin;r_e)
a_{\ell m\lambda}^{\mathrm{src}\dagger}(\wkin;r_e)\ket{0},
\label{sourceWavePacket}
\ee
with the support of $f_{\ell m\lambda}$ restricted by Eq.~\eqref{boundedLocal}. Here $\ell\geq1$, $-\ell\leq m\leq\ell$, and $\lambda$ labels the two physical photon polarizations.
This state is outgoing relative to the black hole and later incoming relative to the localized scattering process. It is neither identified with a mode originating on $\Iminus$ nor with the complete state on $\Hplus$.

For the transmitted component, schematically, exterior spin-one evolution gives
\be
a_{\ell m\lambda}^{\mathrm{out}}(\wkin)
=\mathcal T_{\ell}^{(1)}(\wkin)\cdot a_{\ell m\lambda}^{\mathrm{src}}(\wkin;r_e),
\qquad
\Gamma_{\ell}^{(1)}(\wkin)
=\left|\mathcal T_{\ell}^{(1)}(\wkin)\right|^2,
\label{greybodyDefinition}
\ee
where $\mathcal T_{\ell}^{(1)}$ is the transmission amplitude and $\Gamma_{\ell}^{(1)}$ is the unit-flux transmission probability. 

The first relation isolates the transmitted branch; a complete canonical input-output transformation would also include the reflected branch.
At the density-matrix level we write the transmitted state as
\be
\widetilde\rho_{\mathrm{tr}}
=\mathcal E_{\mathrm{tr}}[\rho_{\mathrm{src}}].
\label{transmittedState}
\ee
The low-frequency suppression of $\Gamma_{\ell}^{(1)}$ affects the transmitted probability, but it does not alter the frequency relation in Eq.~\eqref{sourceFrequency}.

For comparison with the conditional Hawking realization discussed in Appendix~\ref{DR}, the photon number flux of the full Hawking ensemble at infinity has the standard form \cite{Page1976}
\be
\frac{\dd N_\gamma}{\dd t\,\dd\wkin}
=\frac{1}{2\pi}\sum_{\ell,m,\lambda}
\frac{\Gamma_{\ell}^{(1)}(\wkin)}
{e^{\wkin/T_H}-1}\,.
\label{photonFlux}
\ee
Here $T_H=(8\pi GM)^{-1}$ is the Hawking temperature.
Equation~\eqref{photonFlux} describes the unconditioned flux of the full Hawking ensemble. A flux assigned to a selected conditional sector would additionally contain the corresponding conditioning weight. The sparsity of the full Hawking flux is relevant to any practical event-rate estimate \cite{Gray2016}.

\subsection{Gravitational redshift}

Now, we can consider the process when the prepared radiation travels from the near-horizon region to a fixed exterior region. As demonstrated by Pound and Rebka \cite{Pound1960, Pound1965}, a particle that travels away from a black hole would be redshifted. Consider a particle traveling in a Schwarzschild black hole spacetime, with the metric in retarded coordinates written as
\be\label{SchwarzBH}
\dd s^2=-\left(1-\frac{2GM}{\rho}\right)\dd u^2-2\dd u \dd \rho+2\rho^2\gamma_{z\bar z}\dd z\dd \bar z\,,
\ee
where $\gamma_{z\bar z}=\frac{2}{(1+z\bar z)^2}$ and $(z,\bar z)$ are stereographic coordinates.
For a wave vector $k^\mu$, an observer with unit timelike four-velocity $u^\mu$ measures the frequency
\be
\omega[u]=-k_\mu u^\mu.
\label{observerFrequency}
\ee
Let $\xi^\mu=(\partial_t)^\mu$ denote the stationary Killing vector. The corresponding conserved Killing frequency is
\be
\wkin=-k_\mu\xi^\mu.
\label{killingFrequency}
\ee
At a finite exterior radius $r>r_H$, a static observer follows the Killing flow with four-velocity
\be
u_{\mathrm{stat}}^\mu(r)=N(r)^{-1}\xi^\mu.
\label{staticObserver}
\ee
Combining Eqs.~\eqref{observerFrequency}-\eqref{staticObserver}, the frequency measured by this observer is therefore
\be
\wloc(r)=\frac{\wkin}{N(r)}.
\label{localKilling}
\ee
Since $N(r)$ vanishes at the horizon, the static frame is defined only for $r>r_H$. Accordingly, the limit $r_e\to r_H^+$ should be understood as a sequence of emission states and static observers located at finite exterior radii approaching the horizon.

Now consider a fixed scattering radius $r_s>r_H$. The Killing frequency is conserved during propagation, so the local frequencies measured at $r_e$ and $r_s$ are related by
\be
N(r_e)\cdot\wloc(r_e)
=
N(r_s)\cdot\wloc(r_s).
\label{frequencyRelation}
\ee
Hence, one has
\be
\wloc(r_s)
=
\frac{N(r_e)}{N(r_s)}\cdot\wloc(r_e).
\label{finiteRadiusRedshift}
\ee
For the bounded-local-frequency family introduced in Eq.~\eqref{boundedLocal}, $\wloc(r_e)$ remains finite as $r_e\to r_H^+$. Since $r_s$ is kept fixed and $N(r_e)\to0$, one obtains
\be
\wloc(r_s)\to0
\qquad\text{as}\qquad
r_e\to r_H^+.
\label{finiteRadiusSoftLimit}
\ee
Thus, with the black hole background, the scattering radius $r_s$, and the hard local kinematics held fixed, radiation emitted with bounded local frequency increasingly close to the horizon reaches the fixed exterior scattering region in a controlled soft limit. This is the family of states considered in the following analysis.

\begin{figure*}
\begin{center}
\includegraphics[width=12cm]{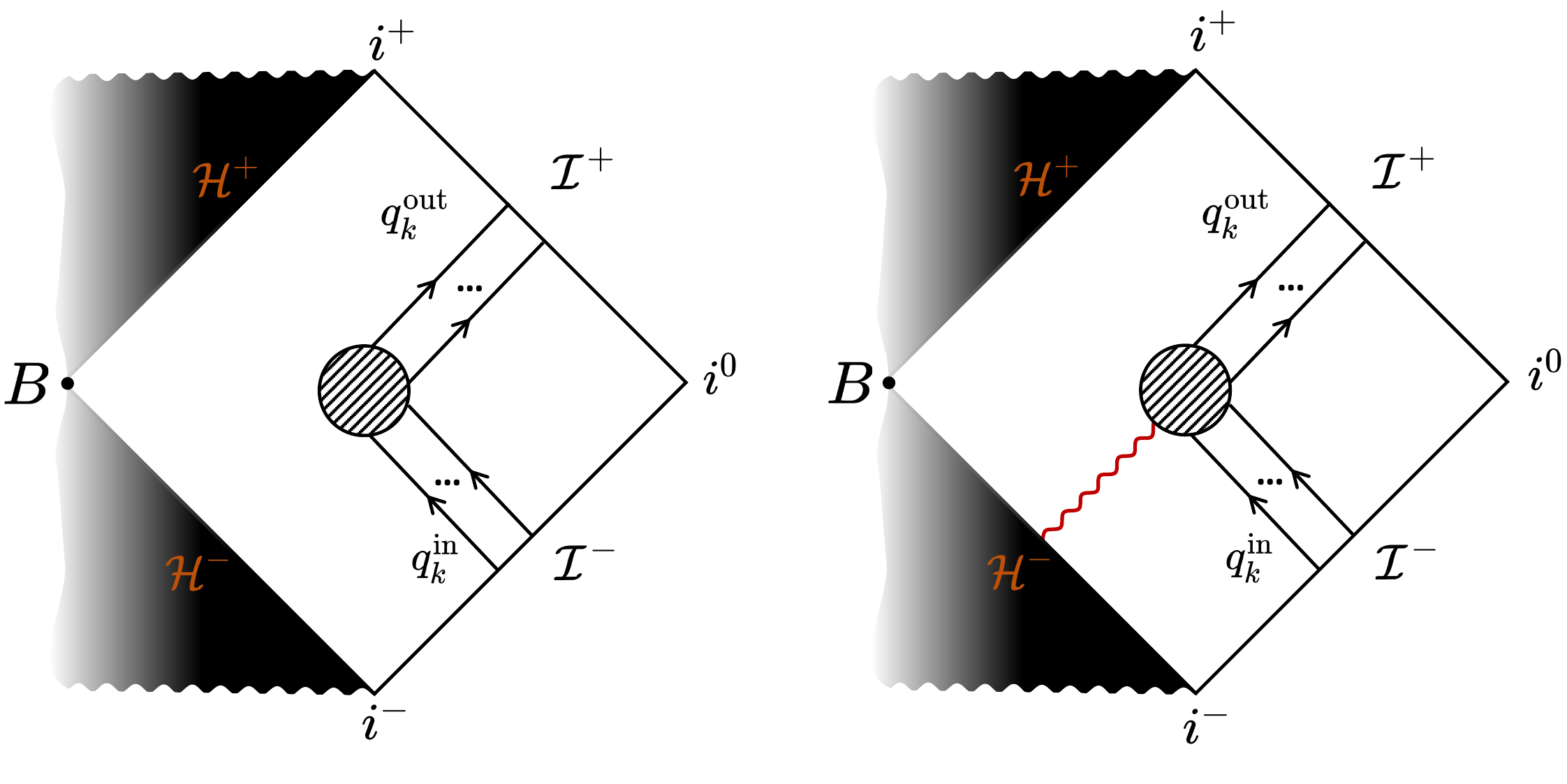}
\end{center}
\caption{
Two schematic scattering processes on an asymptotically flat black hole background. The left panel shows the localized hard scattering process without additional radiation from the near-horizon source. In the right panel, an additional wave packet emitted near the horizon propagates through the spin-one Schwarzschild channel before entering the same hard scattering region. The difference between the two scattering outcomes characterizes the information carried by the transmitted near-horizon radiation.
}
\label{radiation}
\end{figure*}

\section{Finite-distance scattering of the transmitted radiation}
\label{measure}

For the near-horizon radiation considered here, the frequency measured in a fixed exterior region becomes increasingly soft as the emission point approaches the horizon. This makes its direct observation challenging, but it does not mean that its physical effects are inaccessible. A natural alternative is to let the radiation participate in another process and infer its presence from the resulting scattering response. We therefore consider a localized scattering process at a large but finite radius, where the transmitted near-horizon radiation enters as a soft particle. Its leading contribution is then controlled by the universal soft behavior of the scattering amplitude. This simple observation suggests a direct connection between near-horizon radiation, soft theorems, and the infrared triangle in black hole spacetime. In this section, we develop this scattering picture and examine what information can be encoded in the resulting response.

Let us consider scattering the radiation with a set of particles coming from the past null infinity, and finally going to the future null infinity. 
We can consider the following two processes, as shown in Fig. \ref{radiation}, and compare them to see the effect of scattering the transmitted near-horizon wave packet. 
First, we can consider the process without the additional transmitted photon, as shown in the left figure of Fig. \ref{radiation}. We have $n$ incoming hard particles with momenta $q^{\mathrm{in}}_k$ coming from the past null infinity $\mathcal{I}^-$, scattering with each other in the bulk between the horizon and infinity. Then we get $m$ outgoing hard particles with momenta $q^{\mathrm{out}}_j$ going back to the future null infinity $\mathcal{I}^+$. The scattering process happens at a large distance away from the black hole. We can denote the process with S-matrix $\mathcal{S}$ as
\be
\bra{\text{out}}\mathcal{S}\ket{\text{in}}
\ee

Now, we are going to add a massless bosonic particle prepared at $r_e>r_H$ and propagated to $r_s$ to the scattering process, as shown in the right figure of Fig. \ref{radiation}. The particle, specialized below to a photon, is outgoing from the black hole but incoming to the collision and has local tetrad momentum $p_{\mathrm{loc}}^{\hat a}(r_s)$ and polarization $\epsilon^{\hat a}$.
The main task of this section is to answer what the difference is between the two processes shown in Fig. \ref{radiation}.

\subsection{Soft photon theorem}

For the transmitted spin-one state satisfying Eq.~\eqref{finiteRadiusRedshift}, the right panel of Fig.~\ref{radiation} admits a controlled local soft expansion based on Weinberg's soft photon theorem \cite{Weinberg:1965nx}. The calculation is explicitly restricted to the electromagnetic channel defined in Eq.~\eqref{greybodyDefinition}.
For simplicity, we are only going to demonstrate the photon case, while extensions to other spins or gauge sectors require separate propagation and soft-factor analyses.

The collision is described by wave packets supported in an interaction region of size $L_{\mathrm{int}}$ around $r_s$. We impose
\be
L_{\mathrm{int}}\ll\mathcal R_c(r_s),
\qquad
\wloc(r_s)\ll E_{k,\mathrm{loc}}(r_s),
\label{localValidity}
\ee
where $\mathcal R_c(r_s)$ is the local curvature radius. In an orthonormal tetrad,
\be
g_{\mu\nu}e_{\hat a}{}^\mu e_{\hat b}{}^\nu
=\eta_{\hat a\hat b},
\qquad
\eta_{\hat a\hat b}=\operatorname{diag}(-1,+1,+1,+1),
\label{tetradConvention}
\ee
the finite-distance soft parameter is defined by
\be
p_{\mathrm{loc}}^{\hat a}(r_s)
=\lambda\, \widehat p_{\mathrm{loc}}^{\hat a}(r_s),
\qquad \lambda\to0.
\label{localSoftParameter}
\ee
All contractions below are local tetrad contractions, such as $p_{\hat a}\epsilon^{\hat a}$, with $p_{\hat a}\epsilon^{\hat a}=0$. Thus, $q_k+p$ in the following familiar external-line argument is a momentum in the local approximately Minkowski patch, not a globally conserved Schwarzschild four-momentum.

\begin{figure*}
\begin{center}
\includegraphics[width=12cm]{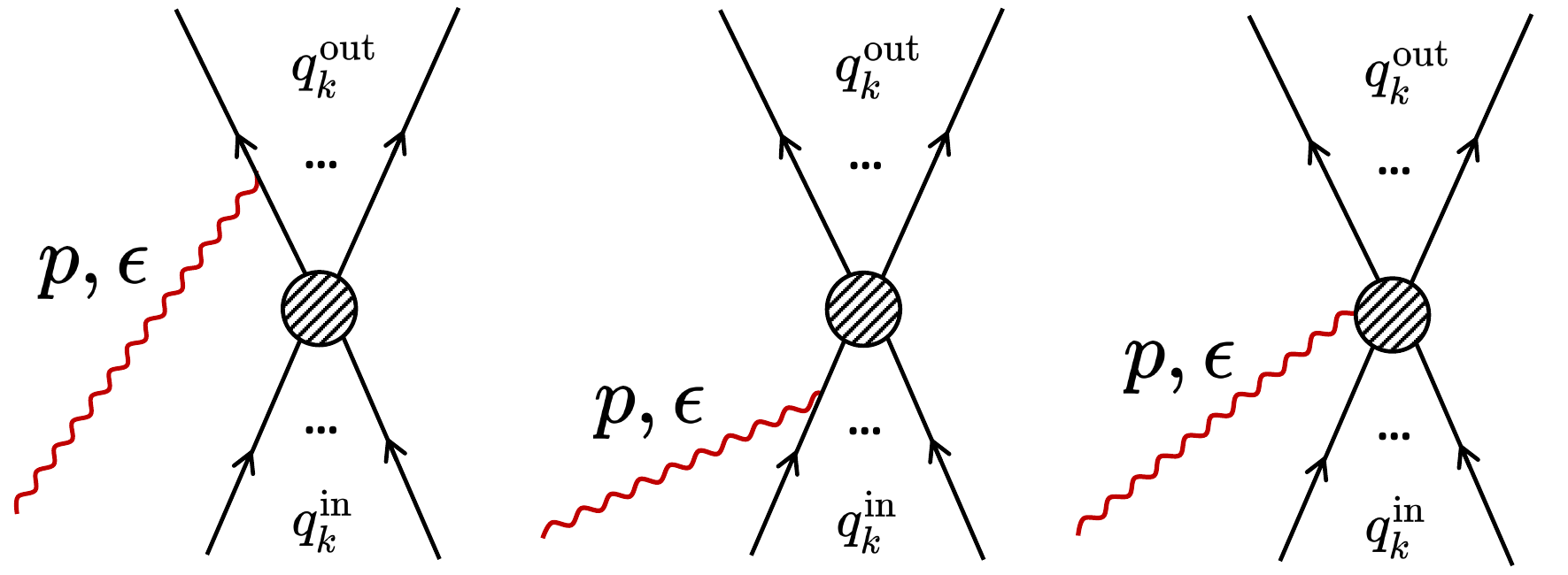}
\end{center}
  \caption{There are three different ways of adding an extra soft photon to the scattering process with $n$ incoming and $m$ outgoing particles. The extra soft photon can be added to the $n$ incoming external lines, $m$ outgoing external lines, or the internal lines.}\label{soft}
\end{figure*}

Now, let us consider adding an extra soft photon with local momentum $p_{\mathrm{loc}}^{\hat a}$ and polarization $\epsilon^{\hat a}$ to a scattering process with $n$ incoming and $m$ outgoing particles, as shown in Fig. \ref{soft}. We can consider the following interaction
\be
\mathcal{L}\sim -A^{\mu}j_{\mu}, ~~~~j_{\mu}=iQ(\Phi\partial_{\mu}\Phi^*-\Phi^*\partial_{\mu}\Phi)\,.\label{int}
\ee
To keep the original derivation readable, $p$ and $q_k$ in Eqs.~\eqref{int}-\eqref{softT} denote the corresponding local tetrad momenta at $r_s$.
First of all, let us attach a soft photon to the incoming external legs, which is illustrated in the first figure of Fig. \ref{soft}. 
Note that we are considering the massless scalar particle $\Phi$ with charge $Q$\footnote{Or one can think that the mass of the particle is taken to be zero afterward since charged particles moving at the speed of light might not be well-defined \cite{Tolish:2014bka}.}.
Throughout this section, $Q$ (or $Q_k$ for the $k$th hard leg) includes the gauge coupling.
From Feynman's rule, the difference between the diagram with and without the extra photon is a propagator with momentum $q^{\text{in}}_k+p$ and a vertex with interaction \eqref{int}. The external legs are all on-shell, so the extra propagator with momentum $q^{\text{in}}+p$ can be expressed as
\be
\frac{-i}{(q^{\text{in}}_k+p)^2+\mu_k^2}=\frac{-i}{2q^{\text{in}}_k \cdot p}\,,
\ee
where we have used $p^2=0$ and $(q^{\text{in}}_k)^2+\mu_k^2=0$. Up to subleading corrections, the vertex can be written as
\be
2i\epsilon^{\mu}Q_k q^{\text{in}}_{k\mu}\,.
\ee
So, the difference between the amplitudes with and without the attachment of an extra soft photon to the external incoming legs is a factor
\be
\frac{Q_k q^{\text{in}}_k\cdot \epsilon}{q^{\text{in}}_k\cdot p}\,.\label{in}
\ee 
The case when the soft photon is attached to an outgoing external leg, shown in the second figure of Fig. \ref{soft} ~is similar, and we also get an extra factor containing a propagator and a vertex
\be
-\frac{Q_k q^{\text{out}}_k\cdot \epsilon}{q^{\text{out}}_k\cdot p}\,.\label{out}
\ee
Note that the above factors \eqref{in} and \eqref{out} are both divergent as $p\to 0$. 
Attachments to internal lines are subleading provided that the hard kinematics remain nonexceptional. In particular, each internal propagator must stay a finite distance away from its mass shell in the soft limit. Thus, for every internal momentum $Q_I$, there exists a constant $\delta_I>0$, independent of the soft momentum $p$, such that
\be
\left|Q_I^2+m_I^2\right|>\delta_I
\qquad\text{as}\qquad p\to0.
\label{nonexceptional}
\ee
If instead an internal line approaches its mass shell in this limit, it develops an additional factorization channel and must be treated separately.

Conditioning on the transmitted one-photon wave packet entering the collision as an incoming soft particle, the leading soft relation takes the form
\bea
\bra{\mathrm{out}}\mathcal S \,a^{\dagger}(p,\epsilon)\ket{\mathrm{in}}
&=&
\left(\sum_{k=1}^{n}Q_k
\frac{q_k^{\mathrm{in}}\cdot\epsilon}
     {q_k^{\mathrm{in}}\cdot p}-
\sum_{j=1}^{m} Q_j \frac{q_j^{\mathrm{out}}\cdot\epsilon} {q_j^{\mathrm{out}}\cdot p}\right)\bra{\mathrm{out}}\mathcal S\ket{\mathrm{in}}+\cdots,
\label{softT}
\eea
where $\cdots$ denotes subleading terms in the soft expansion. Here $n$ and $m$ denote the numbers of incoming and outgoing hard particles, respectively.
Note that there is a $+1$ sign assigned to incoming hard legs and a $-1$ to outgoing hard legs. Since the near-horizon photon is incoming to the scattering process, the leading soft factor is
\begin{equation}
S_{\mathrm{in}}^{(0)}(p,\epsilon)
=\left[
\sum_{k=1}^{n}Q_k^{\mathrm{in}}
\frac{q_k^{\mathrm{in}}\cdot\epsilon}
{q_k^{\mathrm{in}}\cdot p}
-\sum_{j=1}^{m}Q_j^{\mathrm{out}}
\frac{q_j^{\mathrm{out}}\cdot\epsilon}
{q_j^{\mathrm{out}}\cdot p}
\right].
\end{equation}
Replacing $\epsilon^{\mu}$ by $p^{\mu}$ gives
\begin{equation}
S_{\mathrm{in}}^{(0)}(p,p)
=\left[
\sum_{k=1}^{n}Q_k^{\mathrm{in}}
-\sum_{j=1}^{m}Q_j^{\mathrm{out}}
\right]
=0.
\end{equation}
Here the last equality follows directly from charge conservation.
The transmitted photon is incoming to the localized collision, while the corresponding outgoing-soft relation is obtained by crossing.
The key ingredient of understanding the physics of the soft theorem 
is to transform the soft factor to position space using the relationship between a null vector in the bulk and a point labeled by the stereographic coordinates $(z,\bar z)$ on the celestial sphere and horizon, which was worked out in \cite{Cheng:2022xyr, Cheng:2022xgm}.


In the isotropic coordinates $(t,x^i)$, the metric \eqref{SchwarzBH} can be rewritten as
\be\label{isoBH}
\dd s^2=-\left(\frac{r-r_h}{r+r_h}\right)^2\dd t^2+\left(\frac{r+r_h}{r}\right)^4\dd\vec x^2\,,
\ee
with $r_h={GM}/{2}$. 
Here $r$ is the isotropic radius, distinct from the areal radius $\rho$ in Eq.~\eqref{SchwarzBH}; in this coordinate only, the lapse is $N(r)=(r-r_h)/(r+r_h)$. Thus, one needs to notice the difference between $r_h=GM/2$ and the areal-radius horizon $r_H=2GM$.
The coordinate transformation is
\be\label{trans}
t=u+\rho+2GM\ln \frac{\rho-2GM}{2GM}\,, \quad \left(1+\frac{GM}{2r}\right)^2r=\rho\,,
\ee
and
\be
x^1=r\frac{(z+\bar z)}{1+z\bar z},\quad x^2=r\frac{-i(z-\bar z)}{1+z\bar z},\quad x^3=r\frac{1-z\bar z}{1+z\bar z}\,.
\ee
Note that $r$ is the isotropic radial coordinate, which is defined in \eqref{trans}.

At the scattering radius $r_s$, it is convenient to introduce a local orthonormal frame and use the same stereographic coordinates to parametrize null directions. The incoming soft momentum can be written as
\be
p^{\hat a}=\frac{\wloc(r_s)}{1+z\bar z}\left(1+z\bar z,z+\bar z,-i(z-\bar z),1-z\bar z\right).
\ee
For positive helicity, the polarization vector may be chosen as
\be
\epsilon^{+\hat a}=\frac{1}{\sqrt{2}}\left(\bar z,1,-i,-\bar z \right).
\ee
The opposite-helicity polarization vector is obtained similarly. This local parametrization gives the standard celestial representation of the leading soft factor.

Using the stereographic directions of the local tetrad momenta, and absorbing the gauge coupling into $Q_k$ in the celestial expression, the angular soft kernel can be written as
\begin{equation}
\begin{split}
&\sum_{k=1}^n\frac{Q_k q^{\text{in}}_k\cdot \epsilon}{q^{\text{in}}_k\cdot p}
-\sum_{k=1}^m\frac{Q_k q^{\text{out}}_k\cdot \epsilon}{q^{\text{out}}_k\cdot p}
\\
&=
\frac{1+z\bar z}{\sqrt{2}\wloc(r_s)}
\times\left[\sum_{k=1}^{n} \frac{Q^{\textrm{in}}_k }{z-z_k^{\textrm{in}}}
-\sum_{j=1}^{m} \frac{Q^{\textrm{out}}_j }{z-z_j^{\textrm{out}}}\right]\,.
\end{split}
\label{large}
\end{equation}
With the help of the isotropic coordinates \eqref{isoBH}, we are able to rewrite everything in a more compact formulation than the ones in \cite{Cheng:2022xyr, Cheng:2022xgm}. 
Substituting \eqref{large} into the soft theorem, we have 
\begin{equation}
\begin{split}
&\bra{\mathrm{out}}\mathcal{S} \, a^{\dagger}(p,\epsilon)\ket{\mathrm{in}} \\
&=\frac{1+z\bar z}{\sqrt{2}\,\wloc(r_s)}
 \times
 \left[ \sum_{k=1}^{n} \frac{Q^{\mathrm{in}}_k}{z - z_k^{\mathrm{in}}}
   - \sum_{j=1}^{m} \frac{Q^{\mathrm{out}}_j}{z - z_j^{\mathrm{out}}}
 \right]
 \bra{\mathrm{out}}\mathcal S \ket{\mathrm{in}}
 +\cdots\,.
\end{split}
\label{softT2}
\end{equation}

The frequency appearing in Eq.~\eqref{softT2} is measured locally at the scattering point. 
The original radial prefactor is instead the conversion between the Killing frequency and the energy measured locally at the scattering point
\be
\frac{N(r_s)}{\wkin}=\frac{1}{\wloc(r_s)}.
\label{radialConversion}
\ee
Thus, the frequency-dependent prefactor can also be written as
\be
\frac{1+z\bar z}{\sqrt{2}\wloc(r_s)}
=\frac{1+z\bar z}{\sqrt{2}\wkin}
\cdot
\frac{r_s-r_h}{r_s+r_h}.
\label{radialConversionIsotropic}
\ee
This relation shows explicitly how the black hole redshift connects the local celestial soft factor with the conserved Killing frequency. At fixed $r_s$, the limit $r_e\to r_H^+$ drives $\wloc(r_s)$ to the soft regime through Eq.~\eqref{finiteRadiusRedshift}. For $r_s \gg r_h$, the redshift factor approaches unity and the standard asymptotically flat celestial expression is recovered.

The charge-conservation check above makes the gauge-invariance content of the leading soft factor explicit. At finite $r_s$, Eq.~\eqref{softT2} gives the corresponding local soft factorization in an orthonormal frame. In the asymptotic limit, the same leading soft relation is equivalently expressed as the Ward identity associated with large gauge transformations. After the scattered radiation propagates to $\Iplus$, the memory observable constructed below supplies the zero-frequency radiative record and completes the connection among the three corners of the infrared triangle.

\subsection{Infrared-safe inclusive probability and memory effect}
\label{irsafe}

The leading $1/\wloc(r_s)$ scaling is an amplitude-level statement, not by itself a prediction for a detection rate or energy flux. We first combine the hard state with the propagated source state
\be
\widetilde\rho_{\mathrm{tr}}
=\mathcal E_{\mathrm{tr}}[\rho_{\mathrm{src}}],
\qquad
\rho_{\mathrm{in}}
=\rho_{\mathrm{hard}}\otimes\widetilde\rho_{\mathrm{tr}}.
\label{inclusiveInput}
\ee
For a resolved hard outcome represented by $\Pi_{\mathcal D}$ and energy resolution $\Delta E$, an infrared-safe inclusive probability has the form
\be
P_{\mathcal D}(\Delta E)
=\sum_{\substack{X_{\mathrm{unres}}\\E_{X_{\mathrm{unres}}}<\Delta E}}
\operatorname{Tr}\!\left[
\left(\Pi_{\mathcal D}\otimes
\ket{X_{\mathrm{unres}}}\bra{X_{\mathrm{unres}}}\right)
\mathcal S\rho_{\mathrm{in}}\mathcal S^\dagger
\right].
\label{inclusiveProbability}
\ee
The additional unresolved real-soft states are combined order by order with the corresponding virtual-soft corrections. Equivalently, the observable may be formulated using appropriately dressed asymptotic states \cite{Kapec2017,Gabai2016}. In this way the soft scaling and the low-frequency greybody suppression enter the same probability rather than being interpreted separately.

A numerical event rate would also require a source flux, target density, wave-packet overlap, detector response, and signal-to-noise analysis. These inputs are beyond the scope of this work. 
Equation~\eqref{inclusiveProbability} provides an infrared-safe scattering observable for the transmitted state. A concrete event rate can be obtained once the source flux, target density, wave-packet overlap, detector response, and signal-to-noise characteristics are specified. The memory operator defined below provides a complementary asymptotic probe of the same radiation process.

\begin{figure}
\begin{center}
\includegraphics[width=5cm]{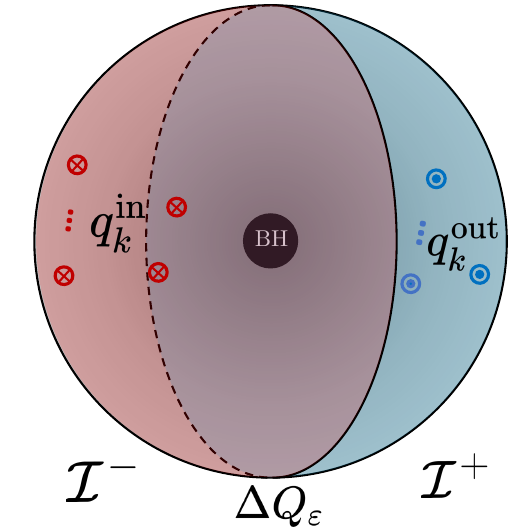}
\end{center}
\caption{An illustration of the 2-dimensional celestial sphere, where we have welded two celestial spheres at $\mathcal{I}^-$ and $\mathcal{I}^+$ together. The scattering process in the bulk can be represented as correlators on the celestial sphere.  We inject $n$ incoming particles with momentum $q_k^{\text{in}}$ to the bulk and receive $m$ outgoing particles with momentum $q_k^{\text{out}}$. For the conditioned process, the early-to-late change of the smeared asymptotic charge is denoted by $\Delta Q_{\varepsilon}$.}
\label{CS}
\end{figure}

Now, we have rewritten the soft factor in terms of the local scattering data and their angular labels. The question is how to properly understand the above soft theorem.
Consider an observer located at infinity sending in $n$ incoming particles with momentum $q_k^{\text{in}}$ to the bulk and receiving $m$ outgoing particles with momentum $q_k^{\text{out}}$, as shown in Fig. \ref{CS}. When the transmitted near-horizon photon joins the scattering process, the observer would find an extra factor with a pole as shown in \eqref{softT2}.
The pole in Eq.~\eqref{softT2} organizes the universal low-frequency response of the amplitude. Its observable content enters the infrared-safe inclusive probability after unresolved real and virtual contributions are combined, while its conditioned zero-frequency angular component is recorded by electromagnetic memory after propagation to $\Iplus$.

After the collision at $r_s$, the scattered radiation is propagated to future null infinity. Electromagnetic memory is then defined by the retarded-time integral of the leading radiative field \cite{Bieri2013},
\be
\widehat{\Delta A}_A(z,\bar z)
=\int_{-\infty}^{+\infty}\dd u\,
\widehat F^{(0)}_{uA}(u,z,\bar z),
\label{memoryOperator}
\ee
and a smooth function $\varepsilon$ selects an angular projection,
\be
\widehat M[\varepsilon]
=\int_{S^2}\dd^2z\,\sqrt\gamma\,
D^A\varepsilon(z,\bar z)\widehat{\Delta A}_A(z,\bar z).
\label{smearedMemory}
\ee
For a phase-invariant thermal density matrix, the unconditioned one-point function vanishes in the absence of a coherent phase. A classical memory profile is therefore associated only with an appropriate coherent or classical limit. For an individual quantum event, the relevant object is instead the expectation value or distribution of Eq.~\eqref{smearedMemory} in a state conditioned on the accompanying hard-scattering record. The process order is preparation at $r_e$, exterior propagation to $r_s$, finite-distance scattering, propagation to $\Iplus$, and finally the retarded-time integration in Eq.~\eqref{memoryOperator}.

Following the proposal in \cite{Strominger:2014pwa, Mao:2017wvx, Pasterski:2015zua, Pasterski:2015tva}, the soft factor \eqref{large} can be related to the zero-frequency radiative field in an appropriate coherent or classical limit. 
At leading order, if we represent the scattering amplitude as $M_{n+m+1}(q_k,[p;\epsilon^{\pm}])$, the soft factor can be denoted as
\be
S^{\pm}=\lim_{\wkin\to0}\wkin\frac{M_{n+m+1}(q_k,[p;\epsilon^{\pm}])}{M_{n+m}(q_k)}\,.
\ee
The classical field profile below applies in the coherent or classical limit, while individual quantum events are described by the conditioned operator expectation introduced afterward.
Then, the change of fields at the celestial sphere can be obtained and read as
\be
\begin{split}
	\Delta A_z=2\left[\sum_{k=1}^{n} \frac{Q^{\textrm{in}}_k }{z-z_k^{\textrm{in}}}-\sum_{k=1}^{m} \frac{Q^{\textrm{out}}_k }{z-z_k^{\textrm{out}}}\right]\,,\\
\Delta A_{\bar z}=2\left[\sum_{k=1}^{n} \frac{Q^{\textrm{in}}_k }{\bar z-\bar z_k^{\textrm{in}}}-\sum_{k=1}^{m} \frac{Q^{\textrm{out}}_k }{\bar z-\bar z_k^{\textrm{out}}}\right]\,.\label{24}
\end{split}
\ee
Thus, in the coherent or classical limit, an asymptotic observer can associate the conditioned scattering process with the field change in Eq.~\eqref{24}.
The above configuration corresponds to a flat gauge connection
, which can be understood from the point of view of the memory effect.

Let $C_u$ denote a cut of $\Iplus$ at fixed retarded time $u$, with sphere measure $\dd^2z\sqrt\gamma$. For specified early and late cuts $C_{u_1}$ and $C_{u_2}$, the memory charge is
\be
\Delta Q_\varepsilon
=Q_\varepsilon[C_{u_2}]-Q_\varepsilon[C_{u_1}]
=-\int_{S^2}\dd^2z\sqrt\gamma\,
\varepsilon D^A\Delta A_A.
\label{memoryChargeDefinition}
\ee
The celestial delta function is normalized by
\be
\int_{S^2}\dd^2z\sqrt\gamma\,
\frac{\delta^{(2)}(z-z_k)}{\sqrt\gamma}F(z,\bar z)
=F(z_k,\bar z_k).
\label{deltaNormalization}
\ee

Now, the change of fields \eqref{24} corresponds to a net charge change.
With the incoming/outgoing convention, the corresponding memory charge can be written as
\be
\Delta Q_{\varepsilon}
=-8\pi\left[
\sum_{k=1}^{n}Q_k^{\mathrm{in}}
\varepsilon(z_k^{\mathrm{in}},\bar z_k^{\mathrm{in}})
-\sum_{j=1}^{m}Q_j^{\mathrm{out}}
\varepsilon(z_j^{\mathrm{out}},\bar z_j^{\mathrm{out}})
\right].
\label{DQ}
\ee
This follows from
\bea
D^A\Delta A_A
=8\pi\left[
\sum_{k=1}^{n}Q_k^{\mathrm{in}}
\frac{\delta^{(2)}(z-z_k^{\mathrm{in}})}{\sqrt\gamma}
-\sum_{j=1}^{m}Q_j^{\mathrm{out}}
\frac{\delta^{(2)}(z-z_j^{\mathrm{out}})}{\sqrt\gamma}
\right].
\label{memoryDeltaIdentity}
\eea

For a source state $\rho_{\mathrm{src}}$, define the state-dependent output
\be
\rho_{\mathrm{out}}(\rho_{\mathrm{src}})
=
\mathcal S
\left(
\rho_{\mathrm{hard}}
\otimes
\widetilde\rho_{\mathrm{tr}}(\rho_{\mathrm{src}})
\right)
\mathcal S^\dagger,
\label{stateDependentOutput}
\ee
where $\widetilde\rho_{\mathrm{tr}}=\mathcal E_{\mathrm{tr}}$ denotes the state after exterior propagation and transmission. If $\Pi_h$ denotes a specified hard-scattering record, the conditioned state is
\be
\rho_{\mathrm{out}|h}(\rho_{\mathrm{src}})=\frac{\Pi_h\rho_{\mathrm{out}}(\rho_{\mathrm{src}})\Pi_h}{\operatorname{Tr}[\Pi_h\rho_{\mathrm{out}}(\rho_{\mathrm{src}})]},
\label{conditionedOutput}
\ee
and the conditioned memory expectations are
\be
\left\{ \operatorname{Tr}\!\left[\widehat M[\varepsilon]\rho_{\mathrm{out}|h}(\rho_{\mathrm{src}})\right]\right\}_{\varepsilon,h}.
\label{sourceMemoryMap}
\ee
The function $\varepsilon$ selects the angular projection being measured, while the label $h$ retains the correlation with the accompanying hard-scattering outcome. The family in Eq.~\eqref{sourceMemoryMap} therefore provides an operational criterion for source distinguishability. Two source preparations can be distinguished through this channel whenever their conditioned memory records differ for at least one choice of $(\varepsilon,h)$. These conditioned expectations indicate which source-dependent features survive exterior propagation and remain accessible at infinity, thereby providing an additional asymptotic constraint or a partial fingerprint of the transmitted radiation. Consistent with the soft-decoupling analyses in Refs.~\cite{Mirbabayi2016,Bousso2017}, the accessible information is naturally encoded in these conditioned asymptotic observables.


\section{Conclusion and Discussion}
\label{Con}

In black hole spacetimes, as in ordinary scattering processes, conservation of the charges associated with large gauge transformations constrains radiative dynamics and connects the leading soft response with electromagnetic memory. In this work, we have followed a specified family of outgoing spin-one radiation states from their preparation near the Schwarzschild horizon, through exterior propagation and finite-distance scattering, to an infrared-safe inclusive probability and a conditioned memory observable at $\Iplus$.

The central result follows from gravitational redshift. For states prepared at $r_e>r_H$ with uniformly bounded local-frequency support, the frequency measured at any fixed exterior scattering radius $r_s$ approaches zero as $r_e\to r_H^+$. The transmitted radiation therefore enters a controlled soft regime at the scattering point. The complete Hawking thermal ensemble is broader, but the bounded-local-frequency particles considered here may be viewed as a particular component of Hawking radiation. This sector is driven toward zero Killing frequency as the preparation point approaches the horizon and has received comparatively little attention in previous discussions.

After propagation through the spin-one Schwarzschild channel, the transmitted wave packet can be treated within a localized weak-curvature region provided that $L_{\mathrm{int}}\ll\mathcal R_c(r_s)$ and $\wloc(r_s)\ll E_{k,\mathrm{loc}}(r_s)$. At finite $r_s$, the radial prefactor $N(r_s)/\wkin=1/\wloc(r_s)$ converts the conserved Killing frequency into the frequency measured locally at the scattering point. The standard asymptotic soft theorem is recovered in the limit $r_s\to\infty$. Once the scattered radiation propagates to $\Iplus$, the triangle relation connects the soft theorem and large-gauge charge conservation with the memory operator.

The inclusive probability and the memory observable provide two complementary probes of the radiation process. By combining unresolved real emissions with the corresponding virtual corrections, the inclusive probability gives a well-defined infrared-safe characterization of the scattering process. The memory observable, in turn, records the conditioned zero-frequency angular component of the radiation at $\Iplus$. Together, these observables connect the universal leading soft behavior of the amplitude with physically accessible information carried by the transmitted radiation.

The conditioned expectation values in Eq.~\eqref{sourceMemoryMap} make explicit how information about the near-horizon source can propagate through the Schwarzschild exterior and become encoded in asymptotic memory. The relevant projection of the transmitted state determines the conditioned memory record, while the smearing function $\varepsilon(z,\bar z)$ allows different angular components to be probed.  They provide a direct criterion for source distinguishability. Whenever different source preparations produce distinct transmitted projections or different correlations with the hard process, these differences can appear in the memory record and, in principle, be identified at infinity. Memory can therefore provide an additional asymptotic constraint and a partial fingerprint of the radiation source. More broadly, these observables help determine which features of near-horizon radiation survive exterior propagation and remain accessible at infinity, providing a concrete way to explore how near-horizon information is imprinted on asymptotic observables.

Extending the present construction to the full thermal ensemble would require carefully distinguishing the successive ingredients of Hawking emission, such as near-horizon production, greybody transmission probabilities, and evaporation rate, before performing the appropriate thermal average.
Finite-temperature soft relations may provide a complementary framework for such an extension \cite{Solanki2023}.


\appendix
\section{Relation to Hawking radiation}
\label{DR}

The main result of the paper is independent of the production mechanism and assumes only the bounded-local-frequency family in Eq.~\eqref{boundedLocal}. 
To relate the bounded-local-frequency family to Hawking radiation, we first review the Damour-Ruffini derivation of the thermal factor in the conserved Killing frequency.
We then relate the resulting single-mode occupation number to the
standard photon flux at infinity, and state the additional source assumption under which an outward component in the pair-creation picture may realize the state family used in the main text.

Let us review the Damour-Ruffini approach for deriving Hawking radiation. Consider a scalar field $\Phi$ living on a 4-dimensional Schwarzschild black hole background, with the Klein-Gordon equation
\be\label{Klein-Gordon}
\frac{1}{\sqrt{-g}}\partial_{\mu}\left(\sqrt{-g}g^{\mu\nu} \partial_{\nu}\Phi \right)-\mu_\Phi^2\Phi=0\,.
\ee
The scalar field $\Phi$ can be separated as
\be
\Phi=\frac{1}{(4\pi\omega)^{1/2}}\frac{1}{r}R_{\omega}(r,t)\,Y_{\ell m}(\theta,\phi)\,,
\ee
where $Y_{\ell m}$ denotes a spherical harmonic.
Working in the tortoise coordinates $r_*$ and in the horizon limit $r\to r_H$, the radial equation of \eqref{Klein-Gordon} can be approximated as
\be
\left(-\partial_t^2+\partial_{r_{*}}^2\right)R_{\omega}(r,t)=0\,.
\ee
In this appendix, $\omega$ in the mode functions is the conserved Killing frequency $\wkin$, rather than a local frequency measured on the horizon.
In the advanced Eddington-Finkelstein coordinates, the incoming and outgoing solutions can be obtained and written as
\bea
R^{\mathrm{in}}_{\omega} &=& e^{-i\wkin v}\,,\\
R^{\mathrm{out}}_{\omega} &=& e^{2i\wkin r}e^{-i\wkin v}
\left(\frac{r-2GM}{2GM}\right)^{i4GM\wkin}\,.
\eea
The outgoing wave cannot be straightforwardly extended to the interior of the horizon because it is singular on the horizon. According to Damour and Ruffini \cite{Damour:1976jd}, we can analytically continue the outgoing wave function on the complex plane and rewrite the outgoing wave function. The part outside of the horizon can be written as
\be
R^{\mathrm{out}}_{\omega}=e^{2i\wkin r}e^{-i\wkin v}
\left(\frac{r-2GM}{2GM}\right)^{i4GM\wkin}.
\ee
The part inside the horizon with consistent normalization is

\be
R^{\mathrm{out}}_{\omega}=e^{4\pi GM\wkin}e^{2i\wkin r}e^{-i\wkin v}
\left(\frac{2GM-r}{2GM}\right)^{i4GM\wkin}.
\ee
The analytic continuation compares the amplitudes of mode branches across the horizon. The resulting relative probability factor is
\be
P_{\omega}=e^{-\beta_H\wkin},
\label{hawkingTemperature}
\ee
with the inverse temperature $\beta_H=8\pi GM$.
$P_{\omega}$ can be regarded as the relative probability factor of creating a pair of particles from vacuum fluctuation.
For Bosonic particles, if we denote $C_\omega$ as the normalization of the zero-particle sector, the constraint that the overall probability of creating zero, one, two, $\cdots$ pairs of particles is 1 can be written as
\be
\sum_{n=0}^{\infty}C_{\omega}P^{n}_{\omega}=1\,.
\ee
The above constraint gives us
\be
C_\omega=1-P_\omega\,.
\ee
Then the mean occupation of particles with energy $\omega$ created at the horizon can be expressed as
\be
\langle n_\omega\rangle
=(1-P_\omega)\sum_{n=0}^{\infty}nP_\omega^n
=\frac{P_\omega}{1-P_\omega}
=\frac{1}{e^{\beta_H\wkin}-1}.
\label{correctOccupation}
\ee
The above result gives the Bose-Einstein occupation number for a single bosonic mode at the Hawking temperature \cite{Hawking1975, Hawking:1974rv}. Although the preceding derivation was presented for a scalar field, the same thermal factor applies to each photon mode. After propagation through the Schwarzschild potential, each spin-one partial wave reaching infinity is weighted by the greybody factor $\Gamma_{\ell}^{(1)}(\wkin)$. 
With the standard continuum normalization, the number of modes in a
frequency interval $\dd\wkin$ during a time interval $\dd t$ is
$\dd t\,\dd\wkin/(2\pi)$.
Summing over angular modes and polarizations gives the photon number flux
 \begin{equation}
\frac{\dd N_\gamma}{\dd t\,\dd\wkin}
=\frac{1}{2\pi}
\sum_{\ell,m,\lambda}\langle n^{\mathrm{out}}_\omega\rangle
=\frac{1}{2\pi}\sum_{\ell,m,\lambda}
\frac{\Gamma_{\ell}^{(1)}(\wkin)}
{e^{\wkin/T_H}-1}\,.
\end{equation}
 The corresponding fermionic result follows from the Fermi-Dirac occupation factor.

Note that the same result can be derived from the so-called Unruh approach \cite{Unruh:1976db}. The near-horizon region of a Schwarzschild black hole can be approximated by the Rindler spacetime. The Unruh effect of the Rindler spacetime can be obtained by tracing out half of the Minkowski vacuum. The resulting density matrix is a thermal density matrix with the Hawking temperature. Local derivations likewise distinguish the short-distance correlations relevant to Hawking radiation from the later asymptotic particle interpretation \cite{FredenhagenHaag1990}.

As a brief summary, the Damour-Ruffini and Unruh descriptions characterize Hawking radiation by a thermal distribution in the asymptotic Killing frequency. High-frequency modes are exponentially suppressed, so at the level of the thermal occupation numbers the largest populations occur in the sector of low Killing frequencies. The Tolman relation $T_{\mathrm{loc}}(r)=T_H/N(r)$ shows that modes with Killing frequency of order $T_H$ have increasingly large local frequencies near the horizon, whereas the family considered here has uniformly bounded local-frequency support \cite{Wald2001}. Although the complete thermal Hawking ensemble is not identical to the bounded-local-frequency family introduced in Eq.~\eqref{boundedLocal}, the latter lies within this highly populated low-frequency part of the Hawking spectrum. As the preparation point approaches the horizon, the corresponding Killing-frequency window is driven further toward the soft end of the spectrum.

The connection with the main text becomes most direct in the pair-creation picture. One may condition on an outward-propagating component produced at $r_e>r_H$ whose local-frequency profile remains bounded as $r_e\to r_H^+$. This selected component realizes the class of near-horizon states considered in the main text. Its subsequent propagation is governed by Eqs.~\eqref{sourceFrequency} and \eqref{finiteRadiusRedshift}, so that at any fixed exterior scattering radius it is driven into the soft regime as the preparation point approaches the horizon. In this sense, our construction isolates a well-defined conditional sector of the Hawking emission process and follows how state-dependent correlations carried by the transmitted component can enter soft scattering and, subsequently, electromagnetic memory at future null infinity.

A large part of the literature on Hawking radiation and the black hole information problem begins with the thermal density matrix or with quantities obtained after the corresponding mode average. The distinction between the full thermal ensemble and the locally bounded outward component may therefore be missed when the analysis is formulated only in terms of the usual asymptotic Hawking observables. In particular, the absence of information tied to an individual source preparation from an ensemble-averaged thermal spectrum does not by itself imply that every conditioned outward component is devoid of state-dependent correlations. Such correlations may cease to be visible once the selected component is combined with all other modes in the full thermal average.

The main text keeps this component separate and follows it through the spin-one Schwarzschild transmission channel in Eq.~\eqref{greybodyDefinition}. This makes it possible to ask whether differences carried by the selected outward component survive greybody propagation, affect the subsequent soft scattering process, and appear in electromagnetic memory at future null infinity. When such differences survive, the memory record can supply an additional asymptotic constraint or a partial fingerprint of the transmitted state. 
In this way, the selected Hawking sector provides a controlled channel through which source-dependent information can remain accessible in asymptotic soft and memory observables.

\section*{Acknowledgements}
This work is supported by the National Natural Science Foundation of China (Grant No. 12405073) and the Natural Science Foundation of Tianjin (Grant No. 25JCQNJC01920).



\begin{thebibliography}{100}

\bibitem{Hawking:2015qqa}
S.~W. Hawking, ``{The Information Paradox for Black Holes},''
\newblock 9, 2015.
\newblock \href{http://arxiv.org/abs/1509.01147}{{\ttfamily arXiv:1509.01147
  [hep-th]}}.

\bibitem{Hawking:2016msc}
S.~W. Hawking, M.~J. Perry, and A.~Strominger, ``Soft {{Hair}} on {{Black
  Holes}},'' \href{https://doi.org/10.1103/PhysRevLett.116.231301}{{\em Phys. Rev.
  Lett.} {\bfseries 116} (2016) 231301},
  \href{http://arxiv.org/abs/1601.00921v1}{{\ttfamily arXiv:1601.00921v1}}.

\bibitem{Strominger:2017aeh}
A.~Strominger, {\em {Black Hole Information Revisited}}.
\newblock 2020.
\newblock \href{http://arxiv.org/abs/1706.07143}{{\ttfamily arXiv:1706.07143
  [hep-th]}}.

\bibitem{Strominger2017}
A.~Strominger, ``{Lectures on the Infrared Structure of Gravity and Gauge
  Theory},'' \href{http://arxiv.org/abs/1703.05448}{{\ttfamily arXiv:1703.05448
  [hep-th]}}.

\bibitem{Pasterski:2020xvn}
S.~Pasterski and H.~Verlinde, ``{HPS} meets {AMPS}: {How} soft hair dissolves
  the firewall,'' \href{http://dx.doi.org/10.1007/JHEP09(2021)099}{{\em JHEP}
  {\bfseries 09} (2021) 099}, \href{http://arxiv.org/abs/2012.03850}{{\ttfamily
  arXiv:2012.03850 [hep-th]}}.

\bibitem{Cheng:2020vzw}
P.~Cheng and Y.~An, ``Soft black hole information paradox: {Page} curve from
  {Maxwell} soft hair of a black hole,''
  \href{http://dx.doi.org/10.1103/PhysRevD.103.126020}{{\em Phys. Rev. D}
  {\bfseries 103} no.~12, (2021) 126020},
  \href{http://arxiv.org/abs/2012.14864}{{\ttfamily arXiv:2012.14864
  [hep-th]}}.

\bibitem{Cheng:2021gdr}
P.~Cheng, ``Evaporating black holes and late-stage loss of soft hair,''
  \href{http://dx.doi.org/10.1103/PhysRevD.106.L061904}{{\em Phys. Rev. D}
  {\bfseries 106} no.~6, (2022) L061904},
  \href{http://arxiv.org/abs/2108.10177}{{\ttfamily arXiv:2108.10177
  [hep-th]}}.

\bibitem{Cheng:2023vrx}
P.~Cheng, ``Circumventing the black hole hair-loss problem,''
\href{http://dx.doi.org/10.1103/PhysRevD.108.066014}{{\em Phys. Rev. D}
  {\bfseries 108} no.~6, (2023) 066014},
  \href{http://arxiv.org/abs/2308.08095}{{\ttfamily arXiv:2308.08095
  [hep-th]}}.

\bibitem{Hawking1975}
S.~W. Hawking, ``{Particle Creation by Black Holes},''
  \href{http://dx.doi.org/10.1007/BF02345020}{{\em Commun. Math. Phys.}
  {\bfseries 43} (1975) 199-220}. [Erratum: Commun.Math.Phys. 46, 206 (1976)].

\bibitem{Hawking:1974rv}
S.~W. Hawking, ``Black hole explosions?,''
  \href{http://dx.doi.org/10.1038/248030a0}{{\em Nature} {\bfseries 248}
  no.~5443, (Mar., 1974) 30-31}.

\bibitem{Hawking:1976ra}
S.~W. Hawking, ``Breakdown of predictability in gravitational collapse,''
  \href{http://dx.doi.org/10.1103/physrevd.14.2460}{{\em Phys Rev D} {\bfseries
  14} no.~10, (Nov., 1976) 2460-2473}.

\bibitem{Bondi:1962px}
H.~Bondi, M.~G.~J. van~der Burg, and A.~W.~K. Metzner, ``{Gravitational waves
  in general relativity. 7. Waves from axisymmetric isolated systems},''
  \href{http://dx.doi.org/10.1098/rspa.1962.0161}{{\em Proc. Roy. Soc. Lond. A}
  {\bfseries 269} (1962) 21-52}.

\bibitem{Sachs:1962zza}
R.~Sachs, ``{Asymptotic symmetries in gravitational theory},''
  \href{http://dx.doi.org/10.1103/PhysRev.128.2851}{{\em Phys. Rev.} {\bfseries
  128} (1962) 2851-2864}.


\bibitem{Mirbabayi2016}
M.~Mirbabayi and M.~Porrati, ``Shaving off Black Hole Soft Hair,''
  \href{https://doi.org/10.1103/PhysRevLett.117.211301}{{\em Phys. Rev. Lett.}
  {\bfseries 117} (2016) 211301},
  \href{https://arxiv.org/abs/1607.03120}{{\ttfamily arXiv:1607.03120}}.

\bibitem{Bousso2017}
R.~Bousso and M.~Porrati, ``Soft Hair as a Soft Wig,''
  \href{https://doi.org/10.1088/1361-6382/aa8be4}{{\em Class. Quantum Grav.}
  {\bfseries 34} (2017) 204001},
  \href{https://arxiv.org/abs/1706.00436}{{\ttfamily arXiv:1706.00436}}.


\bibitem{Solanki2023}
D.~N. Solanki and S.~Bhattacharjee, ``Soft Theorems and Memory Effects
  at Finite Temperatures,''
  \href{https://doi.org/10.1140/epjc/s10052-023-12335-8}{{\em Eur. Phys. J. C}
  {\bfseries 83} (2023) 1167},
  \href{https://arxiv.org/abs/2308.02445}{{\ttfamily arXiv:2308.02445}}.

\bibitem{Pound1960}
R.~V. Pound and G.~A. Rebka, ``Apparent weight of photons,''
  \href{http://dx.doi.org/10.1103/physrevlett.4.337}{{\em Physical Review
  Letters} {\bfseries 4} no.~7, (Apr, 1960) 337-341}.

\bibitem{Pound1965}
R.~V. Pound and J.~L. Snider, ``Effect of gravity on gamma radiation,''
  \href{http://dx.doi.org/10.1103/physrev.140.b788}{{\em Physical Review}
  {\bfseries 140} no.~3B, (Nov, 1965) B788-B803}.


\bibitem{Strominger:2013jfa}
A.~Strominger, ``On {{BMS Invariance}} of {{Gravitational Scattering}},''
  \href{https://doi.org/10.1007/JHEP07(2014)152}{{\em JHEP}
  {\bfseries 07} (2014) 152}.

\bibitem{Strominger:2013lka}
A.~Strominger, ``Asymptotic {{Symmetries}} of {{Yang-Mills Theory}},''
  \href{https://doi.org/10.1007/JHEP07(2014)151}{{\em JHEP}
  {\bfseries 07} (2014) 151}.

\bibitem{Strominger:2014pwa}
A.~Strominger and A.~Zhiboedov, ``{Gravitational Memory, BMS Supertranslations
  and Soft Theorems},'' \href{http://dx.doi.org/10.1007/JHEP01(2016)086}{{\em
  JHEP} {\bfseries 01} (Nov., 2016) 086},
  \href{http://arxiv.org/abs/1411.5745}{{\ttfamily arXiv:1411.5745 [hep-th]}}.

\bibitem{He:2014laa}
T.~He, V.~Lysov, P.~Mitra, and A.~Strominger, ``{{BMS}} supertranslations and
  {{Weinberg}}'s soft graviton theorem,''
  \href{https://doi.org/10.1007/JHEP05(2015)151}{{\em JHEP}
  {\bfseries 05} (2015) 151}.

\bibitem{Cheng:2022xyr}
P.~Cheng and P.~Mao, ``{Soft theorems in curved spacetime},''
  \href{http://dx.doi.org/10.1103/PhysRevD.106.L081702}{{\em Phys. Rev. D}
  {\bfseries 106} no.~8, (2022) L081702},
  \href{http://arxiv.org/abs/2206.11564}{{\ttfamily arXiv:2206.11564
  [hep-th]}}.

\bibitem{Cheng:2022xgm}
P.~Cheng and P.~Mao, ``{Soft gluon theorems in curved spacetime},''
  \href{http://dx.doi.org/10.1103/PhysRevD.107.065010}{{\em Physical Review D}
  {\bfseries 107} no.~6, (2023) 065010},
  \href{http://arxiv.org/abs/2211.00031}{{\ttfamily arXiv:2211.00031
  [hep-th]}}.


\bibitem{Mao:2023rca}
P.~Mao and K.~Y.~Zhang,
``Soft theorems in de Sitter spacetime,''
\href{http://dx.doi.org/10.1007/JHEP01(2024)044}{{\em JHEP}
  {\bfseries 01} (2024), 044},
  \href{http://arxiv.org/abs/2308.08861}{{\ttfamily arXiv:2308.08861
  [hep-th]}}

\bibitem{Mao:2023dsy}
P.~Mao, K.~Y.~Zhang and B.~Zhou,
``Near horizon linearized gravity and soft theorem,''
  \href{http://dx.doi.org/10.1103/PhysRevD.109.065022}{{\em Physical Review D}
  {\bfseries 109} no.~6, (2024) 065022},
  \href{http://arxiv.org/abs/2311.03773}{{\ttfamily arXiv:2311.03773
  [hep-th]}}.

\bibitem{Wald2001}
R.~M. Wald, ``The Thermodynamics of Black Holes,''
  \href{https://doi.org/10.12942/lrr-2001-6}{{\em Living Rev. Relativ.}
  {\bfseries 4} (2001) 6},
  \href{https://arxiv.org/abs/gr-qc/9912119}{{\ttfamily arXiv:gr-qc/9912119}}.

\bibitem{FredenhagenHaag1990}
K.~Fredenhagen and R.~Haag, ``On the Derivation of Hawking Radiation
  Associated with the Formation of a Black Hole,''
  \href{https://doi.org/10.1007/BF02096757}{{\em Commun. Math. Phys.}
  {\bfseries 127} (1990) 273-284}.

\bibitem{Page1976}
D.~N. Page, ``Particle Emission Rates from a Black Hole: Massless
  Particles from an Uncharged, Nonrotating Hole,''
  \href{https://doi.org/10.1103/PhysRevD.13.198}{{\em Phys. Rev. D}
  {\bfseries 13} (1976) 198-206}.

\bibitem{Gray2016}
F.~Gray, S.~Schuster, A.~Van-Brunt, and M.~Visser, ``The Hawking
  Cascade from a Black Hole Is Extremely Sparse,''
  \href{https://doi.org/10.1088/0264-9381/33/11/115003}{{\em Class. Quantum
  Grav.} {\bfseries 33} (2016) 115003},
  \href{https://arxiv.org/abs/1506.03975}{{\ttfamily arXiv:1506.03975}}.

\bibitem{Weinberg:1965nx}
S.~Weinberg, ``Infrared {{Photons}} and {{Gravitons}},''
  \href{http://dx.doi.org/10.1103/physrev.140.b516}{{\em Physical Review}
  {\bfseries 140} no.~2B, (Oct., 1965) B516-B524}.

\bibitem{Tolish:2014bka}
A.~Tolish and R.~M. Wald, ``{Retarded Fields of Null Particles and the Memory
  Effect},'' \href{http://dx.doi.org/10.1103/PhysRevD.89.064008}{{\em Phys.
  Rev. D} {\bfseries 89} no.~6, (2014) 064008},
  \href{http://arxiv.org/abs/1401.5831}{{\ttfamily arXiv:1401.5831 [gr-qc]}}.


\bibitem{Gabai2016}
B.~Gabai and A.~Sever, ``Large Gauge Symmetries and Asymptotic States
  in QED,''
  \href{https://doi.org/10.1007/JHEP12(2016)095}{{\em JHEP} {\bfseries 12}
  (2016) 095},
  \href{https://arxiv.org/abs/1607.08599}{{\ttfamily arXiv:1607.08599}}.

\bibitem{Kapec2017}
D.~Kapec, M.~Perry, A.-M.~Raclariu, and A.~Strominger, ``Infrared
  Divergences in QED, Revisited,''
  \href{https://doi.org/10.1103/PhysRevD.96.085002}{{\em Phys. Rev. D}
  {\bfseries 96} (2017) 085002},
  \href{https://arxiv.org/abs/1705.04311}{{\ttfamily arXiv:1705.04311}}.


\bibitem{Bieri2013}
L.~Bieri and D.~Garfinkle, ``An Electromagnetic Analog of Gravitational
  Wave Memory,''
  \href{https://doi.org/10.1088/0264-9381/30/19/195009}{{\em Class. Quantum
  Grav.} {\bfseries 30} (2013) 195009},
  \href{https://arxiv.org/abs/1307.5098}{{\ttfamily arXiv:1307.5098}}.



\bibitem{Mao:2017wvx}
P.~Mao and H.~Ouyang, ``{Note on soft theorems and memories in even
  dimensions},'' \href{http://dx.doi.org/10.1016/j.physletb.2017.08.064}{{\em
  Phys. Lett. B} {\bfseries 774} (2017) 715-722},
  \href{http://arxiv.org/abs/1707.07118}{{\ttfamily arXiv:1707.07118
  [hep-th]}}.

\bibitem{Pasterski:2015zua}
S.~Pasterski, ``Asymptotic {{Symmetries}} and {{Electromagnetic Memory}},''
  \href{https://doi.org/10.1007/JHEP09(2017)154}{{\em JHEP}
  {\bfseries 09} (2017) 154},
  \href{http://arxiv.org/abs/1505.00716}{{\ttfamily arXiv:1505.00716}}.

\bibitem{Pasterski:2015tva}
S.~Pasterski, A.~Strominger, and A.~Zhiboedov, ``{New Gravitational
  Memories},'' \href{http://dx.doi.org/10.1007/JHEP12(2016)053}{{\em JHEP}
  {\bfseries 12} (2016) 053}, \href{http://arxiv.org/abs/1502.06120}{{\ttfamily
  arXiv:1502.06120 [hep-th]}}.




\bibitem{Damour:1976jd}
T.~Damour and R.~Ruffini, ``{Black Hole Evaporation in the
  Klein-Sauter-Heisenberg-Euler Formalism},''
  \href{http://dx.doi.org/10.1103/PhysRevD.14.332}{{\em Phys. Rev. D}
  {\bfseries 14} (1976) 332-334}.

\bibitem{Unruh:1976db}
W.~G. Unruh, ``Notes on black hole evaporation,''
  \href{http://dx.doi.org/10.1103/physrevd.14.870}{{\em Phys. Rev. D}
  {\bfseries 14} no.~4, (Aug., 1976) 870-892}.


%
%
%
%
%

\end{thebibliography}

\providecommand{\href}[2]{#2}\begingroup\raggedright\endgroup

\end{document}